# A Tool and Methodology for AC-Stability Analysis of Continuous-Time Closed-Loop Systems

Momchil Milev     Rod Burt
milev_momtchil@ti.com     burt_rod@ti.com

*Abstract*—Presented are a methodology and a DFII-based tool for AC-stability analysis of a wide variety of closed-loop continuous-time (operational amplifiers and other linear circuits). The methodology used allows for easy identification and diagnostics of ac-stability problems including not only main-loop effects but also local-instability loops in current mirrors, bias circuits and emitter or source followers without breaking the loop. The results of the analysis are easy to interpret. Estimated phase margin is readily available. Instability nodes and loops along with their respective oscillation frequencies are immediately identified and mapped to the existing circuit nodes thus offering significant advantages compared to traditional "black-box" methods of stability analysis (Transient Overshoot, Bode and Phase margin plots etc.). The tool for AC-Stability analysis is written in SKILL™ and is fully integrated in DFII™ environment. Its "push-button" graphical user interface (GUI) is easy to use and understand. The tool can be invoked directly from Composer™ schematic and does not require active Analog Artist™ session. The tool is not dependent on the use of a specific fabrication technology or Process Design Kit customization. It requires OCEAN™, Spectre™ and Waveform calculator capabilities to run.

## 1   INTRODUCTION

ALTHOUGH small-signal stability of analog and mixed-mode signal integrated circuits is a fairly old problem and has been studied in many ways both by circuit theory and in every-day practice, it is still a significant source of problems that may render a circuit non-operational under certain conditions. What is presented in this paper is a method[2] and a tool that allow small-signal circuit AC-stability to be evaluated for a continuous-time closed-loop systems without breaking the loop. This is especially useful where breaking the loop is very hard or impossible without affecting circuit's performance or biasing conditions. Moreover, if the methodology is employed by an automated design analysis tool, it can evaluate the stability not only of main-loop effects but also of local loops often present in current mirrors, bias circuits, emitter followers and other circuits that otherwise could go undetected and untested. Such a tool could allow for automatic loop identification and full-circuit stability analysis, which gives better picture of the circuit's sensitive nodes/loops as opposed to black-box phase-margin AC-analysis. In this way, this method offers several advantages over the traditional methods of evaluating small-signal circuit stability as node pulsing during transient analysis and Bode/phase-margin-plots in AC-analysis.

### 1.1   Method's principle

We use a technique that may be viewed as analogous to time-domain analysis[1] of circuit's transfer function response to a unit step-function[1]. Yet, the method differs from the latter significantly in performance and scope. The technique excites selected or all circuit nodes consecutively by applying an AC-current signal source to the tested node without changing the circuit under inspection at all. Then by frequency-domain analysis of circuit's AC-response it evaluates each node's sensitivity/stability over a broad range of frequencies. Besides the advantages already mentioned, this approach significantly speeds up the simulation compared to time-domain analysis and broadens the range of frequency coverage.

### 1.2   Assumptions and theory behind the problem

It is assumed that the system response can be adequately described by a second-order system transfer function[1] with both real and complex roots - that is a set of real or complex poles and zeros. The complex poles that can cause the system to oscillate are referred to as *dominant poles/roots*. They determine the circuit's *natural (oscillation) frequency*. In an unstable loop, inherent device noise or any signal at this

M. M. Milev is with High Performance Analog division, High Speed business unit at Texas Instruments Inc., Tucson, AZ 85749 USA (e-mail: milev_momtchil@ti.com).

Rod Burt is with High Performance Analog division, Precision Linear Products at Texas Instruments Inc., Tucson, AZ 85749 USA (e-mail: burt_rod@ti.com).

---

[1] also known as "node pulsing"

frequency can start oscillations that lead to overall system instability. While in simulation such conditions are, generally, very difficult to simulate, an unstable system may begin to oscillate quite easily in the real-world. The *natural frequency* and *damping factor* of those oscillations are determined by the dominant root at this frequency, and thus not simply by the magnitude of the transfer function. That is why through means of AC-current excitation (at a node of the supposedly unstable loop) it is possible to determine the natural frequency and damping ratio by a simple Example Stability Plot with a performance index of –43.1 magnitude at 10.471MHz (natural frequency) pseudo-code of the macro for the JG instruction node (within the loop) without disrupting the normal circuit operation. This translates into a quantitative measurement of this loop's stability (*performance index*).

### 1.3 Method Limitations

Oscillations that are induced by large-signal effects (as signal delays due to transistor saturation, transient charge etc) will not be detected by this method. Therefore the method should be applied to continuous-time systems, or systems that at a given point in time could be viewed as continuous-time systems.

## 2 METHOD IMPLEMENTATION

As already mentioned in the introduction, to obtain a quantitative measure of each circuit node's *stability*, we carry out a number of simulations applying an AC-current stimulus with a wide frequency range to every node of the circuit followed by a measurement of AC-circuit response at the same node. Consistent with the assumptions made in 1.2, a *dominant root* at a normalized *natural frequency* ($\omega_n$=1) can be described by a second-order system transfer function:

$$T(s) = \frac{1}{s^2 + 2s\zeta + 1} \quad (1.1)$$

for the magnitude of the complex part we have ($s = i\omega$):

$$|T(\omega)| = \left| \frac{1}{-\omega^2 + 2(i\omega)\zeta + 1} \right| \quad (1.2)$$

To define the *stability plot* we first take the derivative of the magnitude of the system response with respect to frequency and normalize to both to frequency and magnitude. We further take one more time the derivative of the result with respect to frequency and normalize again with respect to frequency also:

$$P(\omega) = \left( \frac{d}{d\omega} \frac{\left(\frac{d}{d\omega}|T(\omega)|\right).\omega}{|T(\omega)|} \right).\omega \quad (1.3)$$

Through the second-order differentiation and normalization, this procedure filters out the effects of the real poles and zeros, while responding to the complex poles and zeros in the system. In this way, this function's plot will produce a negative peak at the *natural frequency* for every complex pole and a positive peak for every complex zero[2]. An example of a *stability plot* is shown on Fig. 4, considered in more detail in section 3.

Furthermore, at the *natural frequency* $\omega = \omega_n = 1$ we have:

$$P(\omega_n) = -\frac{1}{\zeta^2} \quad (1.4)$$

By virtue of (1.4), having measured *stability plot* peak at the *natural frequency i.e.* loop's *performance index* $P(\omega_n)$, we determine loop's *damping ratio* $\zeta$, and according to Table 1 – loop's corresponding phase margin [1]

**Table 1 Key performance characteristics of a second order system or its dominant root.**

| | Time domain | Frequency domain | | Stability |
|---|---|---|---|---|
| $\zeta$ | Percent overshoot [%] | Phase margin[Deg] | Max magnitude | Performance index |
| 1.0 | 0 | - | - | -1.0 |
| 0.9 | 0 | - | - | -1.2 |
| 0.8 | 2 | - | - | -1.6 |
| 0.7 | 5 | 70 | 1.01 | -2.0 |
| 0.6 | 10 | 60 | 1.04 | -2.8 |
| 0.5 | 16 | 50 | 1.15 | -4.0 |
| 0.4 | 25 | 40 | 1.4 | -6.3 |
| 0.3 | 37 | 30 | 1.8 | -11 |
| 0.2 | 53 | 20 | 2.6 | -25 |
| 0.1 | 73 | 10 | 5.0 | -100 |
| 0.0 | 100 | 0 | $\infty$ | $-\infty$ |

## 3 EXPERIMENTAL RESULTS

To demonstrate the virtues of the suggested method, we carried out a number of circuit simulations of which we include an example. For this example, we draw conclusions using a traditional approach (transient overshoot and phase margin plots) next. We compare these results and conclusions with the results produced by the suggested method of stability analysis using the stability plot (1.3) last.

Let's consider a simple 2MHz op-amp circuit shown on Fig. 1. At nominal values of *rzero*, *cload* and *C1*, the gain/phase plots on Fig. 3 show approximate phase margin of 20 degrees. Similarly, from the transient step response we measure approximately 50% overshoot - Fig. 2 . After running the tool and obtaining the stability plot at the output node (shown on Fig. 4), we see that the negative peak of the plot is at 3.2 MHz and its magnitude is about -29. From Table 1, and (1.4) we can approximate the phase margin at slightly below 20 degrees, which in turn corresponds to about 53% overshoot and damping factor of about 0.2. From the gain/phase margins on Fig. 3 can also be inferred that the natural frequency of the loop's oscillations should the phase/gain margins drop to 0 is expected to be between 2.4 MHz (0db gain crossover frequency) and 3.5 MHz (180 degrees phase lag). The latter observation is consistent with the Stability Plot's peak at 3.16 MHz.

---

[2] Complex zeros (positive peaks in this plot) do not directly affect systems stability. In few cases though, it is important to consider the relative position of a complex zero with respect to a close complex pole to determine the significance of the complex pole on the overall system stability.

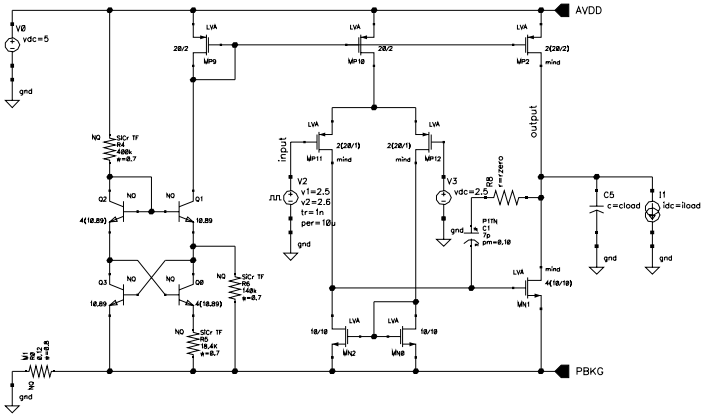

**Fig. 1 Simple 2 MHz op-amp circuit (connected as a buffer)**

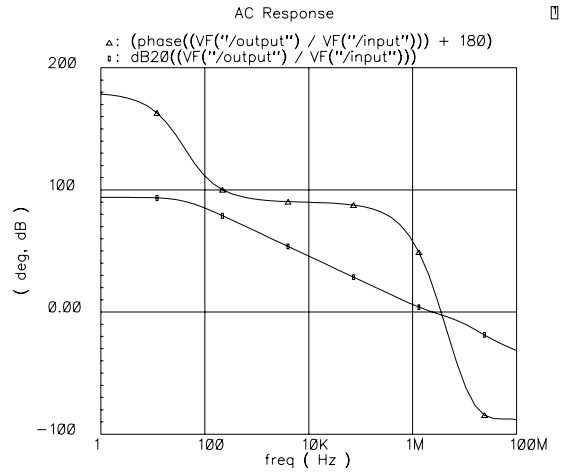

**Fig. 3 Gain/Phase Plot of circuit's AC response showing approximate phase**

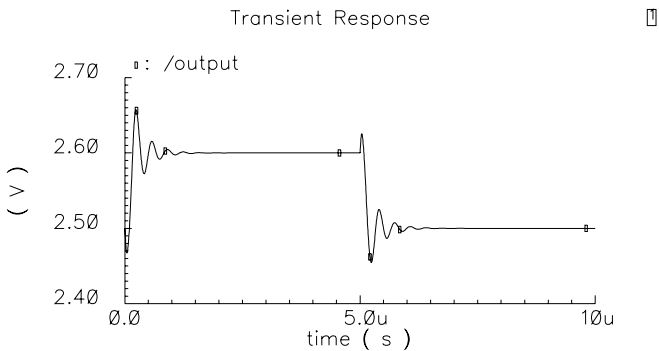

**Fig. 2 Step response showing a corresponding 55% overshoot close to the predicted 53% based on the**

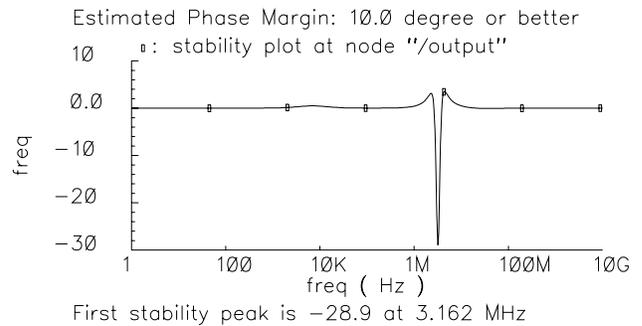

**Fig. 4 Stability Peak at about 3.2 MHz. Magnitude of -28.9 corresponds to about**

While it was relatively easy to obtain the open-loop gain/phase plots for the circuit by opening the main feedback loop, the method proves especially useful in cases where opening the loop would be difficult, or a feedback loop is left "unverified".

To identify all circuit feedback loops and to verify the circuit against possibly unstable loops we applied the method (Section 2) and computed the Stability Plot (negative) peaks at every node of the op-amp circuit. By inspecting the resulting printed report in Table 2, besides the main loop at 3.3 MHz, we identified a local loop inside the zero-TC bias circuit (Figure 5) with natural frequency of about 50 MHz. From Table 1, we inferred that for the nodes of this local loop the equivalent transient step overshoot would range in between 16 and 25% with corresponding phase margin of less than 50 degrees. This helped us realize we needed to compensate this local loop also (by adding a 1pF capacitor at the collector of Q3 for example).

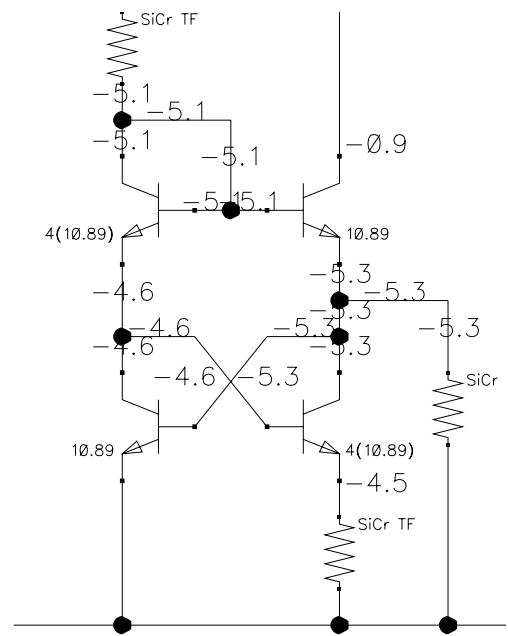

**Figure 5 Bias circuit annotated with Stability Plot values at each node**

**Table 2 Stability Plot peak values for all circuit nodes sorted by loop's natural frequency.**

| Node | Stability Peak | Natural Frequency, Hz |
|---|---|---|
| Loop at 3.3 MHz | | |
| Output | 28.884067 | 3.16E+06 |
| net052 | 28.884063 | 3.16E+06 |
| net136 | 28.884748 | 3.16E+06 |
| net138 | 27.522194 | 3.16E+06 |
| net99 | 27.086771 | 3.31E+06 |
| Loop at 36.3 MHz | | |
| net066 | 0.948229 | 3.63E+07 |
| Loop at 47.9 MHz | | |
| net81 | 5.334409 | 4.79E+07 |
| net17 | 0.504486 | 4.68E+07 |
| net056 | 4.608340 | 4.79E+07 |
| Loop at 51.3 MHz | | |
| net013 | 5.063032 | 4.90E+07 |
| net57 | 4.485003 | 5.01E+07 |
| net16 | 0.252345 | 5.01E+07 |
| net75 | 5.072788 | 4.90E+07 |
| net019 | 0.232893 | 5.13E+07 |

Therefore, the suggested method of using the stability plot to analyze circuit's stability without breaking the main feedback loop proved also very useful in identifying local feedback loops which may require compensation as well.

## 4 THE STABILITY ANALYSIS TOOL

We have implemented a DFII based tool to carry out the task of running a number of circuit simulations determining each circuit node's *stability peak* value, more precisely – loop's *performance index* and *natural frequency* by the method described in sections 1 and 2. The tool is integrated with Analog Artist™ simulation environment through OCEAN™'s application programming interface (API) functions. The tool presently supports Spectre™ and TIspice[3] circuit simulators, but tool's open and modular programming approach easily allows for use of other circuit simulators (Eldo™, cdsSpice™ etc.).

### 4.1 Tool Features

At present, the following features are fully implemented:
- "Single Node" run mode - computes/simulates the stability peak and natural frequency of a single (selected on schematic) node (net). Generates stability peak plot and computes estimated phase margin
- "All Nodes" run mode - computes stability peaks and natural frequencies for all nodes in a circuit/sub-circuit.
- Automatic & Manual Model Setup - auto-configures simulation device model files (if existing environment setup is present), or allows for manual setup/configuration.
- Design Variables Support - existing design variables are imported and configured through a GUI
- "All Nodes" run report - a sorted by each node's natural frequency text report is generated.
- Stability Peak's Special Cases Identification - the "All Nodes" report has been recently augmented with notices alerting the user of special cases: "end-of-range" and "min/max" peak types.
- Analog Artists' scale environment variable support.
- Annotation of Results on circuit schematic.
- Automatic Error and Diagnostic Reporting - auto-generated e-mails will be sent to help in error resolution and tool support.
- Auto-zero all AC sources / stimuli in design prior to running the analysis
- Save and restore original Analog Artist result directory settings*

### 4.2 Features in development

The following features are in various stages of implementation:
- In-tool corners setup
- In-tool sweeps (TEMP etc)
- Remote simulation/distributed/computer farm run capability

## 5 TOOL'S ARCHITECTURE

The tool programming structure is benefiting from modularized code architecture and existing application programming interface (API) functions to interface with the CAD environment of Design Framework II™ (DFII by Cadence Design Systems Inc.). The latter approach allowed us to write code that is tool-independent as much as possible providing for future functionality expansion and support for different circuit simulators. The tool is programmed entirely in SKILL™ utilizing OCEAN™ and Analog Artist's API calls to control DFII's Simulation Environment™ (SE) and a target circuit simulator (Spectre™). Although the tool uses resources generally controlled through Analog Artist's interface, active Analog Artist™ session is not required for the tool to run. The simulation environment setup, simulation job control and simulation results processing is done through OCEAN procedural calls. The simulation task itself is carried by a circuit simulator ("Specter" or other). Generalized tool architecture is shown on Figure 6.

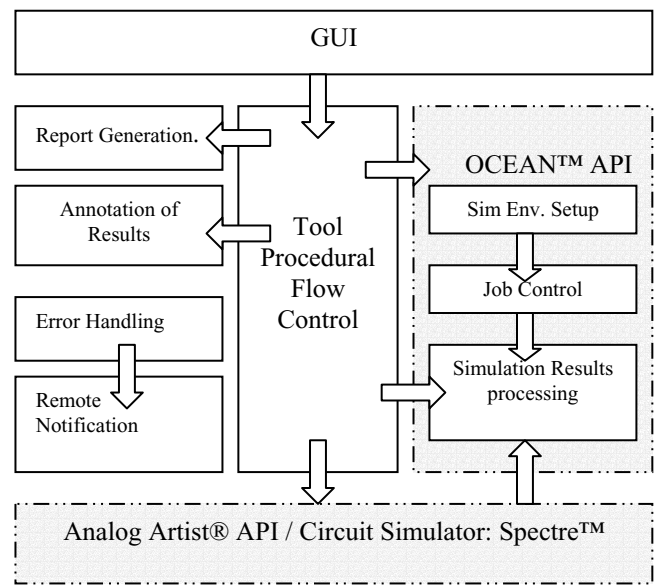

**Figure 6 Stability Analysis Tool architecture**

[3] Tool proprietary to Texas Instruments Inc

## 6 PROGRAM FLOW CONTROL

Tool's procedural control starts with the user selecting either a single-node run mode or all-nodes run mode. In the case of a single-node analysis, an AC-current stimulus source is automatically attached to the net/node selected by the user on the schematic this and an AC-simulation is run across a broad frequency range. The small-signal amplitude of the response is obtained from the simulation results, and the stability plot function (1.3) is used to create the *stability plot* and to estimate the phase margin based on (1.4) and [1].

In both a single-node and all-nodes analysis runs it is challenging to obtain most of the simulation setup parameters (including design variables) automatically from a "current" Analog Artist session. Because there may me more than one active Analog Artist sessions, the auto-configuration of the simulation settings and options is not always trivial. At present current Analog Artist session is considered to be the session referred to by the session-ID returned from asiGetCurrentSession() call. In the future, it is planned to offer a user a way to browse and select from not only his currently active Analog Artist sessions, but also to be able to choose a previously saved Analog Artist's "state" and load most of the simulation setup from there. Due to inconveniences in obtaining the input argument (sevSession-ID) for most of the sev-preffixed procedural calls (sevSaveState(), sevLoadState() etc.) these functions proved not very useful. At future time, when the tool is to be integrated fully under Composer/Analog Artist's GUI these functions will be used and their usage will simplify many of the tasks that need more complex implementation at present.

## 7 CONCLUSIONS AND FUTURE DEVELOPMENT

We showed that the assumption of a second-order system to describe the dominant root is quite adequate in most cases. It provided us with a valuable insight of the system's behavior in analyzing its local and main-loop effects. We developed a tool that proved very useful in determining circuit's feedback loops along with their natural frequencies and damping ratios. A number of features are being added to this tool at present: remote server simulation and distributed computer farm run control, in-tool corner simulation, in-tool DC-sweep (TEMP, device parameters) simulation, importing configuration from Analog Artist's state files and others. Nevertheless, even with the functionality that is offered at present, the tool proved to be very useful in the troubleshooting and analysis of AC-stability problems in a wide variety of linear circuits. The advantages of the method described combined with the automation of the simulation tasks by the tool are easily evident and encourage further development of the functionality of this tool.

**Rod Burt** has been with Texas Instruments Inc (former Burr-Brown Corp) since 1982. As an Analog IC Design Engineer he has designed and led many successful projects helping to build Burr-Brown's high-precision linear products portfolio.

**Momchil Milev** has been with the High Performance Analog EDA Group, of Texas Instruments Inc (former Burr-Brown Corp) since 1998. He has been developing and delivering Process Design Kits, as well as implementing tools and methodologies for electronic design automation. Since May 2004 he is with the High Speed business unit, High Performance Analog (HPA) division of the company. Interests outside Electronic Design Automation include research in Artificial Neural Networks applications and programming of handheld and embedded application devices.